\documentclass[doublecol,a4paper]{epl2} 
\usepackage{graphicx,float,amsmath,amssymb,amsthm}

\usepackage[english,francais]{babel}

\newfont{\ensmathquatorze}{msbm10 scaled 1400}
\newfont{\ensmathonze}{msbm10 scaled 1100}
\newfont{\ensmathdix}{msbm10}
\newfont{\ensmathneuf}{msbm10 scaled 833}
\newfont{\ensmathhuit}{msbm10 scaled 694}
\newfam\ensmathfam                        
\textfont\ensmathfam=\ensmathonze        
\scriptfont\ensmathfam=\ensmathdix       
\scriptscriptfont\ensmathfam=\ensmathhuit
\def\ensmf{\fam\ensmathfam\ensmathonze}         

\newcommand{\ket}[1]{|\kern.3ex#1\kern.3ex\rangle}
\newcommand{\bra}[1]{\langle\kern.3ex #1 \kern.3ex|}
\newcommand{\mean}[1]{\left\langle #1 \right\rangle} 
\newcommand{\smean}[1]{\langle #1 \rangle} 

\newcommand{\EXP}[1]{{\mbox{\large e}}^{#1}}         

\newcommand{\cotg}{\mathop{\mathrm{cotg}}\nolimits}  

\def\RR{{\ensmf R}}                 

\def\I{{\rm i}}                  
\def\D{{\rm d}}                  

\newcommand{\deriv}[2]{\frac{\mathrm{d}#1}{\mathrm{d}#2}}
\newcommand{\derivp}[2]{\frac{\partial #1}{\partial #2}}

\newcommand\antiddots{\mathinner{\mkern2mu\raise1pt\hbox{.}\mkern2mu
\newline \raise4pt\hbox{.}\mkern2mu\raise7pt\hbox{.}\mkern1mu}}

\title{One-dimensional classical diffusion in a random force field \\
       with weakly concentrated absorbers}

\shorttitle{1D classical diffusion in a random force field with dilute absorbers} 

\author{Christophe Texier\inst{1,2}\thanks{\email{christophe.texier@u-psud.fr}} \and Christian Hagendorf\inst{3}\thanks{\email{hagendor@lpt.ens.fr}}}
\shortauthor{C. Texier and C. Hagendorf}

\institute{                    
  \inst{1} Laboratoire de Physique Th\'eorique et Mod\`eles Statistiques,
              Universit\'e Paris-Sud, CNRS, UMR 8626, F-91405 Orsay cedex, France.\\
  \inst{2} Laboratoire de Physique des Solides, Universit\'e Paris-Sud, 
             CNRS, UMR 8502, F-91405 Orsay cedex, France.\\
  \inst{3} Laboratoire de Physique Th\'eorique de l'\'Ecole Normale Sup\'erieure, CNRS, UMR 8549, 24, rue Lhomond, F-75230 Paris cedex 05, France.
}

\abstract{ A one-dimensional model of classical diffusion in a random force field with
  a weak concentration $\rho$ of absorbers is studied. The force field is
  taken as a Gaussian white noise with $\mean{\phi(x)}=0$ and
  $\mean{\phi(x)\phi(x')}=g\,\delta(x-x')$. 
  Our analysis relies on the relation between the Fokker-Planck operator and 
  a quantum Hamiltonian in which absorption leads to breaking of supersymmetry.
  Using a Lifshits argument, 
  it is shown that the average return probability is a power law
  $\smean{P(x,t|x,0)}\sim{}t^{-\sqrt{2\rho/g}}$ (to be compared with the usual
  Lifshits exponential decay $\exp{-(\rho^2t)^{1/3}}$ in the absence of the
  random force field).
  The localisation properties of the underlying quantum Hamiltonian
  are discussed as well. 
}

\begin{document}

\renewcommand{\labelitemi}{$\bullet$}
\renewcommand{\labelitemii}{$\star$}

\selectlanguage{english}

\maketitle

\section{Introduction}
Classical diffusion in a random force field is encountered in several
physical contexts of statistical physics related to anomalous
diffusion or glassy dynamics due to the presence of disorder.  As
introduced by Sinai in the seminal work \cite{Sin82}, it may be
modelised via a Langevin equation
$\dot{x}(t)=2\,\phi(x(t))+\sqrt2\,\eta(t)$, where $\phi(x)$ is a
{\it quenched} random force field with short-range correlations, and
$\eta(t)$ a Langevin force (a normalised Gaussian white noise).
By now its large-time properties in one dimension are well
understood~: in the absence of a global
drift, 
the random force leads to anomalous diffusion characterised by the
scaling of the distance with time $x(t)\sim\ln^2t$. Normal diffusion
properties are only recovered for a sufficiently large drift, however
an intermediate regime reveals several interesting phases (see
Ref.~\cite{BouComGeoLeD90} for a review).  Many approaches to this
problem have been developed~: a probabilistic
method~\cite{BouComGeoLeD90} (a continuous version of the
Dyson-Schmidt method~\cite{LifGrePas88}), Berezinskii diagrammatic
techniques~\cite{GogMel77,Gog82} as well the replica
method~\cite{BouComGeoLeD90}. Moreover more recently, interesting
features of this model like aging properties were analyzed by means of
Ma-Dasgupta real-space renormalisation group
methods~\cite{LeDMonFis99}.

In this letter, we extend the analysis to random media containing randomly 
spread absorbers. Our approach relies on the Fokker-Planck equation (FPE)
\begin{align}
 \label{FPE}
  \derivp{}{t}P
  =\left(\derivp{^2}{x^2}-2\derivp{}{x}\phi(x)-A(x)\right)P
  \equiv -H_\mathrm{FP}P
  \:,
\end{align}
describing classical diffusion in a force field $\phi(x)$ in the
presence of an absorber density $A(x)$.  $P$ denotes the (conditional)
probability $P(x,t|x',0)$ to find a particle at $x$ at time $t$ which
has started from $x'$ at $t'=0$.  We consider the random force field
to be $\phi(x)$ a Gaussian white noise of zero average
$\mean{\phi(x)}=0$ and $\mean{\phi(x)\phi(x')}=g\,\delta(x-x')$
(throughout the paper, $\smean{\cdots}$ will denote averaging with
respect to the quenched disorder~: force field and absorbers).  $A(x)$
describes absorbers at locations $x_n$ with annihilation rates
$\alpha_n>0$, independently and uniformly distributed for a
concentration $\rho$~; thus we write $A(x) = \sum_n \alpha_n \,
\delta(x-x_n)$.  The effects of the random force field and the random
local annihilation rates are well known when described separately. Let
us first review known results for the average return probability
($x=x'$).

\noindent({\it i}) \mathversion{bold}{\it $g=0$ \&
  $\alpha=0$~:}\mathversion{normal} In the absence of random force
  fields and impurities, the conditional probability reads
  $P(x,t|x,0)=\frac1{\sqrt{4\pi\,t}}$.

\noindent({\it ii}) \mathversion{bold}{\it $g\neq0$ \&
  $\alpha=0$~:}\mathversion{normal} For classical diffusion in a
  random force field (Sinai
  problem)~\cite{Sin82,BouComGeoLeD90,LeDMonFis99}, the decay is
  \begin{equation}
    \label{anomalousdiff}
    \smean{P(x,t|x,0)} \underset{t\to\infty}{\sim}  g\, \ln^{-2}(g^2t)
    \:,
  \end{equation}
  and hence much slower than $1/\sqrt{t}$.  This behaviour is related
  to the aforementioned typical distance $x(t)\sim g^{-1}\ln^2(g^2t)$.

\noindent({\it iii})
  \mathversion{bold}{\it $g=0$ \& $\alpha\neq0$~:}\mathversion{normal}
  For free diffusion with a weak absorber concentration we have
  \begin{equation}
  \label{lifshits}
    \smean{P(x,t|x,0)} \underset{t\to\infty}{\sim} \rho(\rho^2t)^{1/6}\,
    \EXP{-3(\frac\pi2)^{2/3}(\rho^2t)^{1/3}}     
    \:.
  \end{equation}
  The rapid (exponential) decay is mostly explained by the decay 
  of the survival probability $\int \D{}x\,\smean{P(x,t|x',0)}\sim\exp-(\rho^2t)^{1/3}$
  (probability is not conserved in the presence of absorption)
  (see \cite{Luc92} for 
  a review on Lifshits tails).
  As switched on from free diffusion, the random force field strongly
  slows down the decay of the probability from $1/\sqrt{t}$ to
  $1/\ln^2t$, whereas absorbers tend to accelerate the decay from a
  power law to exponential $\exp{-t^{1/3}}$.  The aim of the present
  paper is to study the 
  interplay between the random force field and randomly dropped
  absorbers.

\section{ From FPE to Schr\"odinger equation}
Our analysis relies on the well-known relation between the FPE 
(\ref{FPE}) and the Schr\"odinger equation $-\partial_t\psi=H\psi$ 
for~:
\begin{align}
  \label{eqn:hamiltonian}
  H = -\frac{\D^2}{\D x^2} + \phi(x)^2 + \phi'(x) + A(x) 
     \equiv H_\mathrm{susy} + A(x) 
  \:.
\end{align}
A mapping between the two equations is constructed via a non-unitary 
isospectral transformation $P(x,t)=\psi(x,t)\exp{\int^x\phi(x')\D{x'}}$.
In the case $A=0$, this leads to the Hamiltonian 
$H_\mathrm{susy}=(\frac{\D}{\D x}+\phi)(-\frac{\D}{\D x}+\phi)$.
This factorised, {\it supersymmetric}, structure is responsible for a 
positive spectrum. When $\phi$ is a white noise of zero mean, the spectrum
presents a Dyson singularity at the band edge ($E=0$) 
\cite{GogMel77,OvcEri77,Gog82,BouComGeoLeD90}.
The presence of the absorption $A(x)$ in Hamiltonian (\ref{eqn:hamiltonian}) 
breaks the supersymmetry.
Spectral and localisation properties of $H$ were investigated in 
Ref.~\cite{HagTex08} for the case of $A(x)$ and $\phi(x)$ both being 
white noises. This work focused on the mechanism leading 
to a (stochastic) supersymmetry breaking, giving  rise to a 
lifting of the Dyson singularity 
and a breaking of the delocalisation transition at energy $E=0$.  This
first study is connected to the high density limit $\rho\gg\alpha_n$
of the model studied in the present paper. However
Ref.~\cite{HagTex08} has led to the conclusion that the low density
limit $\rho\ll\alpha_n$ studied below is more 
relevant in the context of classical diffusion.

In order to study diffusion properties via \eqref{eqn:hamiltonian},
let us recall that we may relate the return probability, averaged over
the realisations of the random functions $\phi(x)$ and $A(x)$, to the
Laplace transform of its density of states (DoS)~:
\begin{equation}
  \smean{P(x,t|x,0)} = \int\D E\, \rho(E)\,\EXP{-Et}
  \label{eqn:laplace}
  \:.
\end{equation}
Having this relation in mind, we now construct a Lifshits argument for the DoS
$\rho(E)$ of the Hamiltonian~\eqref{eqn:hamiltonian}.

\subsection{Free diffusion with random absorbers ($g=0$ \& $\alpha\neq0$)}
We first recall the famous Lifshits argument~\cite{Lif65,LifGrePas88}
in the absence of the random force field $\phi\equiv0$. Low-energy
states are due to the formation of large impurity-free regions.  Let
us denote by $\ell$ the distance separating two neighboring
impurities.  For $1/\ell\sim\rho\ll\alpha_n$ they impose on the wave
function to vanish at their location (indeed this holds rigorously for
$\alpha_n\to\infty$) and the interval gives rise to a low-energy state
$E_1\simeq(\pi/\ell)^2$.  In terms of classical diffusion, the
survival probability of a diffusive particle released in such an
interval decays as $\EXP{-\pi^2t/\ell^2}$.  Hence the probability for
a low-energy state is related to the probability of a formation of a
large interval~:
$\mathrm{Proba}[E_1<E]=\mathrm{Proba}[\ell>\pi/\sqrt{E}]$.  Since the
distribution of $\ell$ is $\rho\,\EXP{-\rho\ell}$ we recover the
Lifshits singularity \cite{Lif65,LifGrePas88} for the integrated
density of states (IDoS) per unit length
$N(E)\sim\EXP{-\pi\rho/\sqrt{E}}$.  Using a steepest descent method,
we can relate this low energy behaviour to the large time behaviour of
$\smean{P(x,t|x,0)}$ and recover~eq.~(\ref{lifshits})~\footnote{ This
  picture can be generalised in higher dimensions where the main
  exponential behaviour $N(E)\sim\EXP{-\rho\,E^{-d/2}}$ is due to low
  lying states of energy $E\sim1/L^2$ in large regions of volume $L^d$
  free of impurity associated to probability $\EXP{-\rho\,L^d}$~
  \cite{Lif65,LifGrePas88}).  This leads to
  $\smean{P(x,t|x,0)}\sim\exp{-\rho^{\frac{2}{d+2}}t^{\frac{d}{d+2}}}$.
  The preexponential factor of the IDoS has been studied by instanton
  techniques~\cite{FriLut75}.  }.

Let us work out a more precise argument~: in the limit 
 of high impurity weights, 
$\alpha_n\to\infty$, an interval of length $\ell_i$
between two impurities yields a contribution
$\mathcal{N}_0(E;\ell_i)=\sum_{n=1}^\infty\theta(E-(n\pi/\ell_i)^2)$
to the IDoS~:
\begin{equation}
  N(E) \underset{\rho\to0}{\simeq} \lim_{M\to\infty} 
  \frac{\sum_{i=1}^M\mathcal{N}_0(E;\ell_i)}
       {\sum_{i=1}^M\ell_i}
  = \rho\,\mean{\mathcal{N}_0(E;\ell)}_\ell
   \:,
   \label{eqn:idosav}
\end{equation}
where the average is taken with respect to $\ell$ (this approximation
corresponds to the ``pieces model'' of
Ref.~\cite{GreMolSud83,LifGrePas88}). We have
$N(E)=\rho\sum_{n=1}^\infty\int_{n\pi/\sqrt{E}}^\infty\D\ell\,\rho\,e^{-\rho\ell}$
what yields~\cite{BycDyk66a}
\begin{equation}
  \label{singlifshits}
  N(E) \simeq \frac\rho{ \EXP{\pi\rho/\sqrt{E}}-1 }
 \:
 \hspace{0.5cm}\mbox{for } \sqrt{E},\,\rho\ll\alpha.
\end{equation}

\subsection{Random force field with absorbers   ($g\neq0$ \& $\alpha\neq0$)}
We now apply the same argument to the Hamiltonian
(\ref{eqn:hamiltonian}) in order to obtain the low energy DoS.  Due to
the supersymmetric potential $\phi(x)^2+\phi'(x)$, the energy levels
$E_n$ associated to an interval of length $\ell$ differ from
$(n\pi/\ell)^2$, and rather are distributed according to some
nontrivial laws $W_n(E;\ell)$ obtained in Ref.~\cite{Tex00}.  Similarly to
\eqref{eqn:idosav}, we have
\begin{equation}
  \label{sp2}
  N(E) \underset{\rho\to0}{\simeq} \rho\,
           \smean{ \mathcal{N}_\mathrm{susy}(E;\ell) }_\ell 
  \:,
\end{equation}
where $\mathcal{N}_\mathrm{susy}(E;\ell)$ is the IDoS of
$H_\mathrm{susy}$ on $[0,\ell]$ for Dirichlet boundary conditions. 
We 
use the decomposition over the distributions of eigenvalues
$\mathcal{N}_\mathrm{susy}(E;\ell)=\sum_{n=1}^\infty\int_0^E\D E'\,
W_n(E';\ell)$.  Following ~\cite{Tex00}, in the limit $g\ell\gg1$ and
for $E\ll{}g^2$, we may write the distribution of the $n$-th energy
level as
$W_n(E;\ell)\simeq\ell\,N'_\mathrm{susy}(E)\,\varpi_n(\ell\,N_\mathrm{susy}(E))$,
where
$N_\mathrm{susy}(E)=\lim_{\ell\to\infty}\frac1\ell\mathcal{N}_\mathrm{susy}(E;\ell)
=\frac{2g/\pi^2}{J_0(\sqrt{E}/g)^2+N_0(\sqrt{E}/g)^2}$
\cite{OvcEri77,BouComGeoLeD90} is the IDoS per unit length for an {\it
  infinite} volume~; $J_0(x)$ and $N_0(x)$ are the Bessel functions of
first and second kind, respectively. Contrary to
$\mathcal{N}_\mathrm{susy}(E;\ell)$, the IDoS per unit length
$N_\mathrm{susy}(E)$ is insensitive to boundary effects.  Finally
\begin{align}
  N(E)  \underset{\rho\to0}{\simeq} \rho\,\int_0^\infty\hspace{-0.25cm}\D\ell\, 
  \rho\,\EXP{-\rho\ell}
  \sum_{n=1}^\infty\int_0^{\ell N_\mathrm{susy}(E)}
  \hspace{-0.5cm}\D x\,\varpi_n(x)
  \:.
\end{align}
We use the integral representation~\cite{Tex00}
$\varpi_n(x)=\int_\mathcal{B}\frac{\D{}q}{2\I\pi}\,\frac{e^{qx}}{\cosh^{2n}\sqrt{q}}$,
where $\mathcal{B}$ is a Bromwich contour (axis going from
$c-\I\infty$ to $c+\I\infty$ with all singularities of the integrand
at its left), and obtain~:
\begin{align}
   N(E) \simeq \rho\int_0^\infty\hspace{-0.25cm}\D\ell\, \rho\,\EXP{-\rho\ell}
    \sum_{n=1}^\infty
  \int_\mathcal{B}\frac{\D q}{2\I\pi}\,
  \frac{ \EXP{q \ell N_\mathrm{susy}(E)} - 1 }{q\,\cosh^{2n}\sqrt{q}} 
  \:.
\end{align}
If we permute order of integrations, perfom the integral with respect
to $\ell$ and the summation we find
\begin{align}
 \label{Zeintegral}
  N(E) \simeq \rho
    \int_\mathcal{B}\frac{\D q}{2\I\pi}\,
  \frac1{\rho/N_\mathrm{susy}(E) - q}\,
  \frac1{\sinh^{2}\sqrt{q}}
  \:.
\end{align}
Notice that this procedure only converges for a Bromwich contour with
$0<c<q_\ast=\rho/N_\mathrm{susy}(E)$, what always can be attained
via a suitable contour deformation.
Applying the residue theorem to the simple pole at $q_\ast$, we obtain our 
main result for the low energy IDoS per unit length
\begin{equation}
  \label{RES0}
  \boxed{ N(E) \simeq \frac\rho{\sinh^2\sqrt{\rho/N_\mathrm{susy}(E)}} }
  \hspace{0.5cm}\mbox{for }  E \ll g^2,\: \alpha^2
  \:.
\end{equation}
The aforementioned condition $g\ell\gg1$, where $\ell$ denotes the
length of an interval, turns out to be a low density condition
$\rho\ll{}g$ under which \eqref{RES0} is valid.
For completeness, note that $E\gg g^2,\,\alpha^2$ corresponds to the
perturbative regime where we recover the free IDoS
$N(E)\simeq\frac1\pi\sqrt{E}$.
Eq.~\eqref{RES0} allows to identify the energy scale
$E_c=g^2 \EXP{-\sqrt{2g/\rho}}$ separating two regimes.  In the
intermediate energy range, $E_c\ll E \ll g^2$, we recover the IDoS
of~$H_\mathrm{susy}$ 
\begin{equation}
  N(E) \simeq N_\mathrm{susy}(E) \simeq\frac{2g}{\ln^2(g^2/E)}
  \:.
\end{equation}
It is only in the narrow region $0<E\lesssim E_c$ that the scalar
potential $A(x)$ affects the spectrum, for $E\ll{}E_c$~:
\begin{equation}
  \label{RES1}
  \boxed{ N(E)   \underset{\rho\to0}{\simeq} 
  4\rho\,\left(\frac{E}{g^2}\right)^{\sqrt{2\rho/g}} }
  \:.
\end{equation}
Interestingly, we notice that this power law behaviour stems from the
ground state energy $E_1$ of each interval between consecutive
impurities.  Indeed, we may check that eq.~(\ref{RES1}) can be
obtained from
$N(E)\simeq\rho\smean{\theta(E-E_1[\phi(x),\ell])}_{\phi,\,\ell}$, by
means of the distribution $W_1(E;\ell)$, involving
$
\varpi_1(x)=\frac{4}{\sqrt{\pi}\,x^{3/2}}\sum_{m=1}^\infty(-1)^{m+1} m^2\EXP{-m^2/x}
$~\cite{Tex00}. 
{\it A priori} it is far from being obvious that the analysis can be
restricted to the ground state of each interval since the
distributions $W_n(E;\ell)$ have strong overlaps~\cite{Tex00}.

\begin{figure}[!ht]
  \centering
  \includegraphics[scale=0.425]{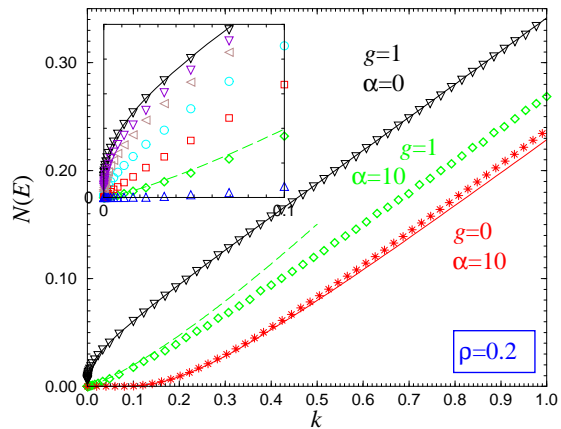}
  \caption{ 
    {\it Comparison between numerical results (triangles, diamonds \&
      stars) and analytical results (lines). 
    }
    Inset~: 
    {\it IDoS for $g=1$ with $\rho$ ranging from $0.01$ to $0.5$.} 
    }
  \label{fig:one}
\end{figure}

Again, combining \eqref{RES1} and a steepest descent argument for
\eqref{eqn:laplace}, we relate the power law behaviour of the IDoS to
a power law decay of the particle density for $t\gg\frac1{g^2}
\EXP{\sqrt{2g/\rho}}$
\begin{equation}
  \label{RES2}
  \boxed{
     \smean{ P(x,t\to\infty|x,0) } \underset{\rho\to0}{\simeq}
     \frac{4\rho\,\Gamma(\sqrt{2\rho/g}+1)}{(g^2t)^{\sqrt{2\rho/g}}}
  }\:.
\end{equation}
The crossover time scale $t_c=g^{-2}\exp{\sqrt{2g/\rho}}$ corresponds
to the time needed by the random particle released in the random force
field to reach the nearest absorber~: $x(t_c)\sim1/\rho$ where
$x(t)\sim{}g^{-1}\ln^2(g^2t)$.  The return probability (\ref{RES2})
decays faster than (\ref{anomalousdiff}) but slower than the
exponential decay (\ref{lifshits}) in the absence of the random force
field.  This provides the answer to our initial question.

\section{Localisation}
We now analyse the localisation properties of the
underlying quantum Hamiltonian~\eqref{eqn:hamiltonian}.

\vspace{0.15cm}

\noindent({\it i}) \mathversion{bold}{\it $g\neq0$ \&
  $\alpha=0$~:}\mathversion{normal}
In the absence of absorption, $A(x)=0$, the 
Lyapunov exponent  (inverse
localisation length) is exactly known~\cite{BouComGeoLeD90} 
$\gamma_\mathrm{susy}(E)=-\frac{g}2k\frac{\D}{\D{}k}\ln[J_0^2(k/g)+N_0^2(k/g)]$.
It reaches a finite value at high energy 
$\gamma_\mathrm{susy}(E\gg{}g^2)\simeq{}g/2$
and vanishes at zero energy~: 
\begin{equation}
  \label{eq:3}
  \gamma_\mathrm{susy}(E\to0)\simeq\frac{g}{\ln(g/k)}\to0
  \:.
\end{equation}

\vspace{0.15cm}

\noindent({\it ii}) \mathversion{bold}{\it $g=0$ \&
  $\alpha\neq0$~:}\mathversion{normal}
On the other hand, in the absence of supersymmetric noise,
$\phi(x)=0$,  and for a low density of 
impurities, $\rho\ll\alpha$, the Lyapunov exponent is given 
by
$\gamma_\mathrm{scalar}(E)\simeq\frac{\rho}{2}\ln[1+(\frac{\alpha}{2k})^2]$
for  $k\gg\rho$ \cite{LifGrePas88,BieTex08}. At high energy
it leads to the well known linear increase of the localisation length
as a function of the energy \cite{AntPasSly81} 
$1/\ell_\mathrm{loc}=\gamma_\mathrm{scalar}(E)\simeq\rho\alpha^2/(8E)$
for $E\to\infty$.
At zero energy it reaches a finite value given by \cite{LifGrePas88,Tex99}
\begin{equation}
  \label{eq:2}
  \gamma_\mathrm{scalar}(E=0)\simeq\rho\,[\ln(\alpha/\rho)-\mathrm{C}]
  \hspace{0.5cm}\mbox{for } \rho\ll\alpha\:,
\end{equation}
where
$\mathrm{C}\simeq0.577$ is the Euler constant. 

\vspace{0.15cm}

\begin{figure}[!ht]
  \centering
  \includegraphics[scale=0.4]{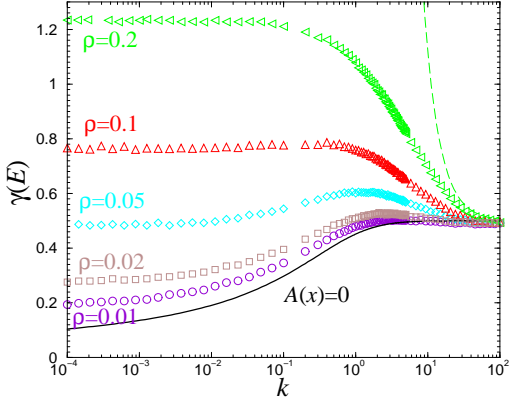}
  \caption{ 
    {\it Low energy Lyapunov exponent as a function of energy for
      $\rho$ ranging from $0.2$ to $0.01$~;
      with $g=1$. Black line~: $\gamma_\mathrm{susy}(E)$.
      Green dashed line~: perturbative result for $\rho=0.2$.
      }
    }
  \label{fig:lyap}
\end{figure}

\noindent({\it iii}) \mathversion{bold}{\it $g\neq0$ \&
  $\alpha\neq0$~:}\mathversion{normal}
We now turn to the case where $\phi$ and $A$ both differ from zero.
The high energy (perturbative) expression is~:
$\gamma(E)\simeq\frac{\rho\alpha^2}{8E}+\frac{g}2$. It corresponds to
the addition of the perturbative expressions for 
$\gamma_\mathrm{scalar}(E)$ and $\gamma_\mathrm{susy}(E)$ (a similar result
was obtained in Ref.~\cite{HagTex08} when $A$ is a Gaussian white noise).
We can see on figure~\ref{fig:lyap} that, for a fixed $g$,
$\gamma(E)$ (obtained numerically) slowly converges 
to $\gamma_\mathrm{susy}(E)$ as $\rho$ is decreased.
The introduction of the scalar potential $A(x)$  
breaks the
delocalisation at $E=0$ obtained for the supersymmetric Hamiltonian.

Let us now turn to the detailed analysis of the zero energy Lyapunov
exponent. 
For this purpose it is convenient to convert the Schr\"odinger
equation $H\psi=E\psi$ into a stochastic differential equation 
for the Ricatti variable 
$z=\psi'/\psi-\phi$~: this latter obeys 
$\deriv{}{x}z=-E-z^2-2z\,\phi(x)+A(x)$. This Langevin like equation
can be related to a Fokker-Planck type equation for the distribution
$\partial_xT(z;x)=\partial_z[(E+z^2)T(z;x)]+2g\partial_z\big[z\partial_z[zT(z;x)]\big]+\rho[T(z-\alpha;x)-T(z;x)]$, where
the first term is a drift term related to the force field
$F(z)=-E-z^2$, the second term a diffusive term and the last term is a
jump term originating from the scalar impurities $A(x)$.
For $E>0$ the Ricatti variable is driven from $+\infty$ to $-\infty$
in a finite ``time'' $x$. The steady current of $z(x)$  corresponds to
its number of divergences per unit length, {\it i.e.}
the number of nodes of the wave function $\psi$ per
unit length, that is to $N(E)$.
The distribution reaches a stationary distribution
$T(z;x\to\infty)=T(z)$  for a steady current~$-N(E)$~:
\begin{equation}
  \label{eq:T}
  N(E) = (E+z^2)T(z) +2gz\deriv{}{z}[zT(z)] 
  - \rho\int_{z-\alpha}^z\hspace{-0.35cm}\D z'\,T(z') 
\end{equation}
The IDoS is given by normalising the solution of this integral equation.
Given $T(z)$,
the IDoS can be extracted from the distribution thanks to the Rice formula
$N(E)=\lim_{z\to\infty}z^2T(z)$. Lyapunov exponent is given by
\cite{LifGrePas88}
$\gamma(E)=\smean{z}=\lim_{R\to\infty}\int_{-R}^{+R}\D{z}\,z\,T(z)$. 

For $E=0$ and positive jumps $\alpha>0$, the Ricatti is constrained to
belong to  $\RR^+$ since both the  ``force'' $F(z)$ and the effect of
the multiplicative noise $\phi(x)$ vanish for $z=0$~: we recover $N(E=0)=0$.
In the low density limit $\rho\to0$ the Ricatti variable is driven to
$z\sim0$ and eventually reinjected at $z\sim\alpha$ with a
``rate'' $\rho$. 
We can write that the negative current due to the ``force'' $F(z)$ and
the noise $\phi(x)$ is equilibrated by the positive current $\rho$ due
to the jumps~:
$\rho\simeq{}z^2T(z) +2gz\deriv{}{z}[zT(z)]$ for $z\in[z_c,\alpha]$.
We have introduced a cutoff $z_c$ where the jump occur almost surely
(the force $F(z)$ and the
multiplicative noise vanish effectively as $z\to0$).
The solution of this equation is 
$
  T(z)\underset{\rho\to0}{\simeq} \frac{\rho}{2g}
  \frac1z\EXP{-z/2g}\int_{z_c}^z\frac{\D z'}{z'}\EXP{z'/2g} 
$. 
This distribution may be approximated by 
$T(z)\simeq\frac{\rho}{2gz}\ln(z/z_c)$ for
$z_c\lesssim{z}\lesssim{g}$ and 
$T(z)\simeq\frac{\rho}{z^2}$ for
$g\lesssim{z}\lesssim\alpha$ (and $T(z)\simeq0$ elsewhere).
Normalisation gives $\ln^2(g/z_c)\simeq\frac{4g}{\rho}$.
Using $\gamma=\mean{z}$, we get
\begin{equation}
  \label{eq:lyapunov}
  \boxed{
  \gamma(0) \underset{\rho\to0}{\simeq} \sqrt{\rho g} + \rho \ln
  (\alpha/g)
  }\:.
\end{equation}
The second term is reminiscent of the result obtained in the absence
of the supersymmetric noise. However the dominant term involves a
nontrivial combination of $\rho$ and $g$.
In the region of rarefaction of eigenstates ($E\sim0$), the localisation
length reads
$\ell_\mathrm{loc}=1/\gamma\sim\frac1\rho\sqrt{\rho/g}\ll1/\rho$.
The scalar impurities breaks the delocalisation transition of the
supersymmetric Hamiltonian, but quite surprisingly the localisation
length is much smaller than the inverse density of impurities.

\section{Numerics}
We now turn to a numerical analysis in order to check our analytical
results for spectrum and localisation 
and explore more precisely their
validity range. 
We analyse the Hamiltonian (\ref{eqn:hamiltonian}) for
$\phi(x)=\sum_n\lambda_n\delta(x-x_n)$ and
$A(x)=\sum_n\alpha_n\delta(x-x_n)$.  This choice of modelisation of the
random force field $\phi$ allows to deal with a continuous
description.  The locations $x_n$ are uniformly distributed
and uncorrelated, with densities $\rho_\phi$ for those of $\phi(x)$
and $\rho$ for impurities of $A(x)$.
The precise shape of the distribution for the dimensionless weights
$\lambda_n$ is not important, as long as it satisfies
$\mean{\lambda_n}=0$ and $\smean{\lambda_n^2}$ finite.  We choose a
symmetric exponential law
$p(\lambda_n)=\frac1{2\lambda}\EXP{-|\lambda_n|/\lambda}$. The process
$\phi(x)$ behaves as a Gaussian white noise in the limit $\rho_\phi\to\infty$
and $\lambda\to0$, with
$g=\rho_\phi\smean{\lambda_n^2}=2\rho_\phi\lambda^2$ fixed. 

\begin{figure}[!ht]
  \centering
  \includegraphics[scale=0.425]{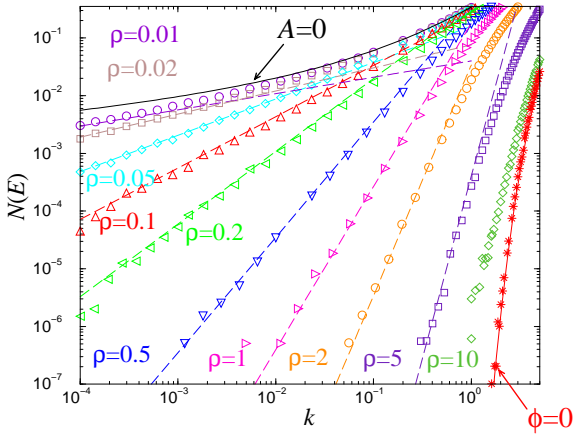}
  \caption{ 
    {\it Dashed lines correspond to
    $N(E)\propto{E}^{\sqrt{2\rho/g}}$.
    Impurity weight is $\alpha=50$ and the number of impurities is
    $N_i=10^7$ 
    ($N_i=10^8$ for the largest densities and smallest energies). 
    The red line is
    eq.~(\ref{singlifshits}) 
    and the red stars are the
    corresponding numerical calculation,  for $\rho=10$.}
    }
  \label{fig:two}
\end{figure}

\subsection{Phase formalism}
Let us now explain how
we can obtain the spectral density from the phase
formalism~\cite{AntPasSly81,LifGrePas88}.  In the equation
$H\psi=E\psi$, we replace the couple of variables $(\psi,\psi')$ by
the couple $(\theta,\xi)$ defined as $\psi=\EXP{\xi}\sin\theta$ and
$\psi'-\phi\psi=k\EXP{\xi}\cos\theta$, where $E=k^2$. The two
variables obey the differential equations~\cite{HagTex08}~:
\begin{eqnarray}
  \label{pf1}
  \frac{\D\theta}{\D x} &=& 
        k - \frac{A(x)}k \sin^2\theta + \phi(x)\,\sin2\theta \\
  \label{pf2}
  \frac{\D\xi}{\D x} &=& 
            \frac{A(x)}{2k} \sin2\theta - \phi(x)\,\cos2\theta
\end{eqnarray}
We solve these equations as follows
\begin{itemize}
\item 
  We denote $\ell_n=x_{n+1}-x_n$. 
  The evolution of the variables on an interval free of impurity is
  $\theta_{n+1}^--\theta_n^+=k\ell_n$ and $\xi_{n+1}^--\xi_n^+=0$, 
  where $\theta_n^\pm\equiv\theta(x_n^\pm)$ and
  $\xi_n^\pm\equiv\xi(x_n^\pm)$ are values just before ($-$) or right
  after ($+$) the $n$-th impurity. The size of these intervals is
  distributed according to the Poisson law
  $P(\ell)=\rho_\mathrm{tot}\EXP{-\rho_\mathrm{tot}\ell}$ for a
  density $\rho_\mathrm{tot}=\rho+\rho_\phi$.

\item
The effect of a $\delta$-peak of $A(x)$ is given by
$\cotg\theta_n^+ - \cotg\theta_n^- = \frac{\alpha_n}{k}$ 
and 
$\xi_n^+   - \xi_n^- = \ln\frac{\sin\theta_n^-}{\sin\theta_n^+}
  =\frac12\ln
    \big[
    1+\frac{\alpha_n}{k}\sin2\theta_n^-+\frac{\alpha_n^2}{k^2}\sin^2\theta_n^-
    \big]
$ (see Ref.~\cite{Tex99}).

\item
The effect of a $\delta$-peak of $\phi(x)$ is given by
$\tan\theta_n^+ = \tan\theta_n^-\,\EXP{2\lambda_n}$ and 
$\xi_n^+   - \xi_n^- =\frac12\ln\frac{\sin2\theta_n^-}{\sin2\theta_n^+}
  = \frac12\ln
    \big[
   \EXP{2\lambda_n}\sin^2\theta_n^-+\EXP{-2\lambda_n}\cos^2\theta_n^-
   \big]$  (see Ref.~\cite{BieTex08}).
\end{itemize}   
It is worth noticing that the two effects on the envelope of the wave
function are roughly given by
$\Delta\xi_n=\xi_n^+-\xi_n^-\sim\ln\alpha_n$ and 
$\Delta\xi_n\sim|\lambda_n|$.

The IDoS corresponds to the number of zeros of the wavefunction of
energy $E$, {\it i.e.} the number of times the cumulative phase $\theta(x)$
coincides with an integer multiple of $\pi$. Therefore, a convenient
way to get the IDoS numerically is~\cite{AntPasSly81,LifGrePas88}
$N(E)=\lim_{x\to\infty}\frac{\theta(x)}{\pi\,x}$. Moreover, the
damping of the envelope is characterised by the Lyapunov exponent
$\gamma(E)=\lim_{x\to\infty}\frac{\xi(x)}{x}$, providing a definition
of the inverse localisation length ($\ell_\mathrm{loc}=1/\gamma$).

\subsection{Ricatti variable}
The equations of the phase formalism are singular in
the limit $E\to0$. In
order to obtain the zero energy Lyapunov exponent, it is
more simple to perform the analysis in term of the Ricatti variable.
Let us denote $z_n^\pm=z(x_n^\pm)$ its values
before/after the impurity $n$.
Between two impurities the evolution is given by
$\arctan\big({z_{n+1}^-}/{k}\big)-\arctan\big({z_{n}^+}/{k}\big)=-k\ell_n$,
{\it i.e.} $1/{z_{n+1}^-}-1/{z_{n}^+}=\ell_n$ for $E=0$.
Through a scalar impurity of $A(x)$ we have 
$z_{n}^+-z_{n}^-=\alpha_n$ and through an impurity of $\phi(x)$ we
have $z_{n}^+= z_{n}^-\,\EXP{-2\lambda_n}$. IDoS and Lyapunov exponent
can be extracted from the stationary distribution of the Ricatti as
explained above.

\subsection{Results}
As a first check, we consider the case with $\phi\equiv 0$~: we
compare numerics for $\lambda=0$ to the Lifshits singularity
(\ref{singlifshits}) (red stars and red line on figure
\ref{fig:one}) and obtain a good agreement (some deviations appear for 
larger $k$ 
since $N(E\gg\alpha^2)\simeq\frac{\sqrt{E}}{\pi}$). 
 Next, we treat the case 
with  $A=0$ 
and check the numerics (black triangles on
figure~\ref{fig:one}) against the analytical expression recalled
above~\cite{OvcEri77,BouComGeoLeD90} (black continuous line)
$N_\mathrm{susy}(E\to0) =
\frac{g/2}{[\ln(2g/k)-\mathrm{C}]^2+\pi^2/4}
+O\big(\frac{k^2}{\ln^2k}\big)$. 
The agreement to the theory shows that our modelisation
of Gaussian white noise $\phi$ is adequate.

\begin{figure}[!ht]
  \centering
  \includegraphics[scale=0.3]{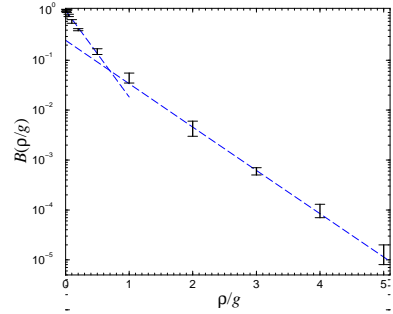}
  \caption{\it Numerical result for function $B(x)$. Lines are purely
    indicative (they correspond to
    $\EXP{-4x}$ and $\EXP{-2x}/4$).}
  \label{fig:BdeX}
\end{figure}

Finally, we combine both noises (green diamonds on
Fig.~\ref{fig:one} and inset), and check against the power law
$N(E\to0)\propto{}E^{\sqrt{2\rho/g}}$~: the exponent fits very well
within a surprisingly 
wide range (see figure~\ref{fig:two} for densities ranging from
$\rho=0.01$ to~$5$). We insist that slopes in the log-log plot
are not fitted but directly compared to~$\sqrt{2\rho/g}$ (straight
dashed lines).
We however observe that, apart for the lowest densities, the prefactor  
significantly differs from $4\rho$, eq.~(\ref{RES1}).
Since the IDoS reaches a finite limit for 
$\alpha\to\infty$, the additional dimensionless factor is a function of
the ratio $\rho/g$ only.
We conclude that  numerics suggests the form
$
  N(E\to0) \simeq 4\rho\,
  B(\rho/g)\,\big(\frac{E}{g^2}\big)^{\sqrt{2\rho/g}}
$ 
with $B(x\to0)=1$. 
The function $B(x)$ is extracted from numerics and plotted on
figure~\ref{fig:BdeX}. 
For $\rho/g=10$,
the numerical precision does not allow a convincing fit for the lowest
energies (green diamonds of Fig.~\ref{fig:two}).
We have however checked that the noise $\phi(x)$ still affects the 
IDoS which has not yet reached
the Lifshits result (\ref{singlifshits}) (red stars and line).

\begin{figure}[!ht]
  \centering
  \includegraphics[scale=0.4]{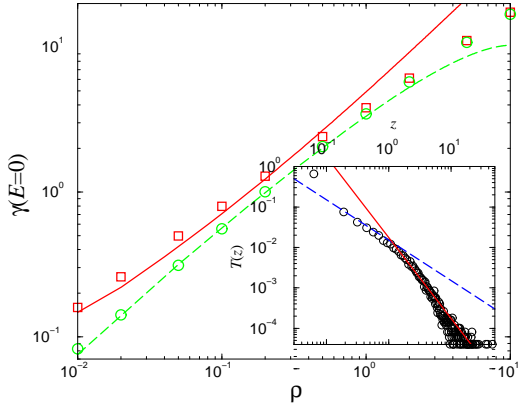}
  \caption{ 
    {\it Lyapunov exponent at $E=0$. Green dashed line~: in the absence of
      supersymmetric noise, for $\alpha=50$. Numerics (circles) is
      compared to equation \eqref{eq:2}.
      Red~: with $\phi(x)$ ($g=1$). 
      Comparison with expression (\ref{eq:lyapunov}).}
      Inset~: {\it Stationary distribution of the Ricatti variable 
      in this latter case
      ($\rho=0.05$)~;
      dashed blue line is $\propto1/z$ and red line $\propto1/z^2$.}
    }
  \label{fig:lyap2}
\end{figure}

The energy dependence of the Lyapunov exponent is obtained from the phase
formalism. The results are plotted for different densities $\rho$ on
figure~\ref{fig:lyap}.
In a second step we analyze the dynamic of the Ricatti variable for 
$E=0$. The stationary distribution is plotted on the inset of
figure~\ref{fig:lyap2} on which we can check the crossover between the
behaviours $T(z)\propto1/z$ and $T(z)\propto1/z^2$ occuring for
$z\sim{g}=1$ (note however that prefactors do not fit with the one
derived above). The distribution is used to compute the zero energy
Lyapunov exponent.  
The analytical result \eqref{eq:lyapunov} is compared to the numerical
result and 
satisfactory coincides (Fig.~\ref{fig:lyap2}).

\section{Conclusion} 
We have studied the average return probability for
one-dimensional classical diffusion in a random force field and the
presence of absorbers at weak concentration ($\rho\to0$), yet with strong
absorption rates.  We have shown that absorption only takes place
above a very large time scale $t_c=g^{-2}\EXP{\sqrt{2g/\rho}}$. 
The well-known Sinai decay
$\smean{P(x,t|x,0)}\simeq{2g}\,{\ln^{-2}(g^2t)}$ holds for $t\ll{t_c}$
and is replaced by the
power law 
$\smean{P(x,t|x,0)}\simeq{4\rho}\,{(g^2t)^{-\sqrt{2\rho/g}}}$
for $t\gg{t_c}$.
Whereas a simple guess would have been to put the crossover between
(\ref{lifshits}) and (\ref{RES2}) at $g\sim\rho$, the power law
(\ref{RES2}) persists numerically up to large ratio $\rho/g$. 
It would be interesting to understand more
carefully the crossover and obtain analytically the prefactor
$B(\rho/g)$ not predicted in our calculation.
Another interesting issue would be to investigate the fluctuations of the
return probability $P(x,t|x,0)$ over disorder configurations~; such a
question is related to the characterisation of the fluctuations of the
local DoS of the quantum 
Hamiltonian, a question studied
for high energies for the scalar noise in
Ref.~\cite{AltPri89} and for the supersymmetric noise in
Ref.~\cite{BunMcK01}.

The localisation properties of the underlying quantum Hamiltonian have
been considered, too. In particular, the scalar impurities lifts the
divergence of the localisation length (inverse Lyapunov exponent) at
energy $E=0$ and leads to a localisation length
$\ell_\mathrm{loc}\sim1/\sqrt{g\rho}$ for~$\rho\to0$.



\end{document}